\documentclass{article}
\usepackage{amsmath}
\usepackage{amssymb}
\usepackage{epsfig}
\usepackage{color}

\makeatletter
\@addtoreset{equation}{section}

\makeatother

\begin{document}

\begin{flushright}
Aug 2010

SNUTP10-007
\end{flushright}

\begin{center}

\vspace{5cm}

{\LARGE 
\begin{center}
On Large $N$ Solution of Gaiotto-Tomasiello Theory
\end{center}
}

\vspace{2cm}

Takao Suyama \footnote{e-mail address : suyama@phya.snu.ac.kr}

\vspace{1cm}

{\it 
BK-21 Frontier Research Physics Division

and 

Center for Theoretical Physics, 

\vspace{2mm}

Seoul National University, 

Seoul 151-747 Korea}

\vspace{3cm}

{\bf Abstract} 

\end{center}

The planar solution is discussed for an ${\cal N}=3$ Chern-Simons-matter theory constructed recently by Gaiotto and Tomasiello. 
The planar resolvent is obtained in terms of contour integrals. 
If the sum of two Chern-Simons levels $k_1,k_2$ is small, the expectation value of a supersymmetric Wilson loop grows 
exponentially with the total 't Hooft coupling, as is expected from AdS/CFT correspondence. 
If one of the Chern-Simons levels, say $k_2$, is taken to infinity, for which 
one of the 't Hooft coupling constants becomes zero, then the exponential behavior disappears.

\newpage

\vspace{1cm}

\section{Introduction}

\vspace{5mm}

Wilson loop operators have played important roles in the study of AdS/CFT correspondence \cite{Maldacena:1997re}. 
It was proposed in \cite{Rey:1998ik}\cite{Maldacena:1998im}\cite{Rey:1998bq} that the expectation value $\langle W \rangle$ 
of a Wilson loop operator in a gauge theory would be equal to the partition function of a 
fundamental string in a certain string theory. 
In the most well-studied example of AdS/CFT correspondence, the gauge theory is ${\cal N}=4$ super Yang-Mills theory, and the 
string theory is Type IIB string theory on AdS$_5\times S^5$. 
A consequence of the proposal is that, in the limit in which the string coupling constant is small and 
the radius of the AdS space is large in the string side, $\langle W \rangle$ is evaluated in terms of 
the area of a minimal surface in the AdS$_5$ space. 
Since the limit corresponds to the large 't Hooft coupling limit in the gauge theory, the result so obtained should be quite 
non-trivial. 

A similar relation between Wilson loops and strings was already proposed long time ago. 
Recently, there was surprising progress on this relation in the context of AdS/CFT correspondence. 
In the case where the gauge theory is ${\cal N}=4$ super Yang-Mills theory, 
a quantitative check of the proposal \cite{Rey:1998ik}\cite{Maldacena:1998im}\cite{Rey:1998bq} 
was performed perturbatively in \cite{Erickson:2000af}, and then the proposal was proved in the above-mentioned limit using the 
localization technique in \cite{Pestun:2007rz}. 
For another case where the gauge theory is ABJM theory \cite{Aharony:2008ug}, 
a similar localization technique for various Chern-Simons-matter theories was developed in 
\cite{Kapustin:2009kz} which can apply to supersymmetric Wilson loops 
constructed in \cite{Drukker:2008zx}\cite{Chen:2008bp}\cite{Rey:2008bh}\cite{Drukker:2009hy}. 
The localization formula obtained in \cite{Kapustin:2009kz} was then studied in the planar limit 
\cite{Suyama:2009pd}\cite{Marino:2009jd}, 
and it was confirmed that the proposal 
was indeed valid for ABJM theory 
\cite{Marino:2009jd}. 
In addition to this result, the planar solution of ABJM theory was further investigated in \cite{Drukker:2010nc} 
and the mysterious $N^{\frac32}$ 
behavior of the free energy, which is another consequence of AdS/CFT correspondence, was derived without relying on calculations 
based on string theory or M-theory. 

\vspace{5mm}

It would be interesting if the investigations mentioned above could be extended to more general superconformal gauge theories. 
The large 't Hooft coupling $\lambda$ behavior of Wilson loops was investigated in \cite{Rey:2010ry} for the planar limit of 
${\cal N}=2$ quiver gauge theory of 
$\hat{A}_1$-type and ${\cal N}=2$ supersymmetric gauge theory with $N_f=2N_c$ fundamental matters. 
The arguments in \cite{Rey:2010ry} were based on the localization of \cite{Pestun:2007rz}. 
In fact, it is quite difficult to analyze the large $\lambda$ limit even in the planar limit. 
Although the localization allows us 
to reduce the problems to the corresponding problems in a matrix model, the matrix model is still complicated for 
${\cal N}=2$ theories due to the 
complicated structure of the 1-loop determinant around the fixed point for the localization. 
Note that the above two theories are related. 
The former theory has the gauge group SU$(N)\times$SU$(N)$. 
Let $g_1,g_2$ be the gauge coupling constants for those gauge fields. 
The latter theory can be obtained (up to a decoupled system) from the former theory 
by taking one of the gauge coupling constant, say $g_2$, to zero. 
See also \cite{Gadde:2009dj}. 
This relation between two theories should descend to a relation for Wilson loops. 

In \cite{Rey:2010ry}, it was argued that a suitable Wilson loop in the quiver gauge theory with $g_1=g_2$ would behave as 
\begin{equation}
\langle W \rangle \sim e^{\sqrt{2\lambda}}
\end{equation}
in the large $\lambda$ limit, 
as expected from AdS/CFT correspondence. 
In the following, we will describe that a Wilson loop has the AdS behavior when the dependence of $\langle W \rangle$ 
on $\lambda$ in the large $\lambda$ limit coincides with the one expected from AdS/CFT correspondence. 
In this terminology, it was argued that a suitable linear combination of Wilson loops in the quiver gauge theory has the 
AdS behavior. 
On the other hand, the Wilson loop in the gauge theory with fundamental matters does not seem to have the AdS behavior. 
This would be 
rather natural from the point of view of the field theory: the Wilson loop would be simply screened by the fundamental matters. 
Note that the contribution of the fundamental matters is not negligible in the planar limit which keeps the conformal symmetry, 
since the number of the fundamental matters should also 
become large in the limit, or one has to take the Veneziano limit, not the 't Hooft limit. 
However, the anticipated non-exponential behavior does not seem to be easily understood from the possible gravity dual point of 
view. 
Since the number of fundamental matters is large, it is not appropriate to realize them in terms of probe D-branes in a similar 
manner as in 
\cite{Gaiotto:2009tk}\cite{Hikida:2009tp}\cite{Ammon:2009wc}\cite{Jafferis:2009th}\cite{Hohenegger:2009as}\cite{Fujita:2009xz}\cite{Benini:2009qs}. 
It was argued in \cite{Rey:2010ry} that a possible resolution might be related to the contributions from 
many, possibly infinite, saddle-points for the string partition function, if they existed. 
All of them were supposed to be 
of the same order of magnitude but might have different phases so that the leading AdS behavior could be canceled 
among them, leaving sub-leading contributions which could be non-exponential. 

Since this observation 
looks quite interesting, and there could be new insights in this kind of situations, it would be worth 
performing similar investigation in more details. 
Due to the complexity of the above example, it would be better to find a simpler setup which may share the same physics. 
The theory to be studied should have a product gauge group, and there should exist a limit in which some of the gauge fields 
decouple. 
If the theory before taking 
the limit has bi-fundamental matters, then there are a large number of fundamental matters after taking 
the limit. 
It would be interesting if $\langle W \rangle$ could be evaluated in such a theory. 

\vspace{5mm}

In this paper, we investigate a Chern-Simons-matter theory constructed recently in \cite{Gaiotto:2009mv}. 
The theory has a product gauge group with two adjustable coupling constants. 
By taking one of the coupling constant to zero, it is reduced to another Chern-Simons-matter theory which contains a large number 
of fundamental matters. 
The localization calculation of \cite{Kapustin:2009kz} can apply to both the original theory and the limit theory. 
We find that the original theory has a range of the coupling constants in which 
a Wilson loop has the AdS behavior, while the corresponding Wilson loop in the limit theory 
behaves at most as a power of the 't Hooft coupling constant. 
Since the localization formula for Chern-Simons-matter theory is much simpler than that for superconformal gauge theories in 
four dimensions, the results mentioned above can be obtained {\it exactly} in the planar limit. 
It should be noted that the investigation of the Chern-Simons-matter theory might not be a good analogy with the system 
considered in \cite{Rey:2010ry} since the gauge fields are non-dynamical in the former. 

\vspace{5mm}

This paper is organized as follows. 
In section \ref{Romans}, we review the construction of \cite{Gaiotto:2009mv} and some properties of their gravity duals. 
Section \ref{largeN} contains the analysis of the planar solution of an ${\cal N}=3$ theory in \cite{Gaiotto:2009mv}. 
The details of the theory is summarized in subsection \ref{GT}, and the planar resolvent of the theory is determined 
in subsection \ref{planar solution}. 
The technique used here is a standard one found, for example, in \cite{Marino:2004eq}. 
The resolvent is given in terms of two contour integrals, supplemented by two equations which 
determine two parameters in the integrals in terms of the coupling constants of the theory. 
Subsection \ref{flavored} shows the planar analysis for the limit theory. 
Section \ref{discuss} is devoted to discussion. 
In Appendix \ref{N=3 action}, we review basic facts on ${\cal N}=3$ Chern-Simons-matter theories. 
For the check of our planar solution obtained in subsection \ref{planar solution}, in Appendix \ref{ABJ} we derive 
the planar resolvent obtained in \cite{Halmagyi:2003ze}, which is related to ABJ theory via a suitable analytic continuation 
\cite{Marino:2009jd}, from our solution. 
A comment on the consistency of the ${\cal N}=3$ theory is made in Appendix \ref{SUSYbreak}.

\vspace{1cm}

\section{Chern-Simons levels and Romans mass} \label{Romans}

\vspace{5mm}

As explained in the introduction, we would like to find a Chern-Simons-matter theory which has a product gauge group, 
and in which one can take a limit such that the gauge fields for one of the simple factors of the gauge group 
decouple. 
One may think that ABJ theory \cite{Aharony:2008gk} would be the right setup. 
Indeed, in ABJ theory, there are two 't Hooft coupling constants 
\begin{equation}
\lambda_1 = \frac{N_1}k, \hspace{5mm} \lambda_2 = \frac{N_2}k. 
\end{equation}
It is possible to take a limit in which all $N_1,N_2,k$ are large while $\lambda_1,\lambda_2$ are kept finite, and then take 
$\lambda_2\to0$ limit. 
Such a limit was recently investigated in \cite{Drukker:2010nc}. 
However, this limit implies $N_1\gg N_2$. 
As a result, the number $N_f$ 
of the fundamental matters, which come from bi-fundamental matters in ABJ theory via the decoupling, 
turns out to be small compared with the rank $N_1$ of the surviving gauge group U$(N_1)$ since $N_f$ is proportional to $N_2$. 
In this case, the effect of the fundamental matters is just a small perturbation, and their effect on the geometry in a possible 
gravity dual would be irrelevant. 

Therefore, we would like to find another theory. 
It turns out that a theory constructed in \cite{Gaiotto:2009mv} is suitable for our purpose. 
The theory has the gauge group U$(N_1)\times$U$(N_2)$, the same group as in ABJ theory\footnote{
Chern-Simons-matter theories with different levels were also discussed in \cite{Fujita:2009kw}. 
}. 
The Chern-Simons levels $k_1,k_2$ 
for those U$(N)$ factors, respectively, do not need to satisfy $k_1+k_2=0$. 
The relevant coupling constants of this theory are 
\begin{equation}
\lambda_1 = \frac{N_1}{k_1}, \hspace{5mm} \lambda_2 = \frac{N_2}{k_2}. 
\end{equation}
In the limit $k_2\to\infty$, the gauge fields for U$(N_2)$ would decouple, and a bi-fundamental matter becomes $N_2$ fundamental 
matters of U$(N_1)$. 
Since $N_2$ can be taken to be of the order of $N_1$, the contribution of the fundamental matters remains relevant even 
after taking the planar limit. 
The price to pay is that the theories in \cite{Gaiotto:2009mv} have at most ${\cal N}=3$ supersymmetry. 
Actually, this is not a bad deal. 
Indeed, ${\cal N}=3$ supersymmetry is powerful enough to write down the Lagrangian of the IR fixed point theory explicitly, and 
more importantly, it makes the localization calculation of \cite{Kapustin:2009kz} available. 
The planar solution of GT theory will be discussed in the next section. 
In the remainder of this section, we will review some properties of the possible gravity duals of those theories argued in 
\cite{Gaiotto:2009mv}. 

\vspace{5mm}

The crucial difference of the theories in \cite{Gaiotto:2009mv} from ABJ theory is that, as mentioned above, 
$k_1+k_2$ is non-zero. 
It was argued in \cite{Gaiotto:2009mv} that $k_1+k_2$ is related to the R-R zero-form field strength $F_0$, or the Romans mass, 
in the dual Type IIA 
string theory. 
In other words, the gravity duals of the theories in \cite{Gaiotto:2009mv} should be massive Type IIA string theories. 

A piece of evidence for this claim is as follows. 
Consider a D0-brane. 
If a non-zero $F_0$ is turned on, then the worldvolume gauge fields $A_\mu$ on the D0-brane have the coupling of the form 
\cite{Bergshoeff:1996cy}\cite{Green:1996bh}
\begin{equation}
\int_{\rm D0} F_0 A. 
\end{equation}
This means that there is a tadpole for $A_\mu$. 
To cancel this tadpole, an appropriate number of fundamental strings should be 
attached to the D0-branes. 
In fact, the number turns out to be $F_0$, so that $F_0$ must be quantized. 
If the background is of an AdS-type, then the opposite ends of those fundamental strings 
should be attached to the boundary of the AdS space. 
AdS/CFT correspondence claims that the dual 
boundary field theory should have a local operator corresponding to the D0-brane with 
$F_0$ fundamental strings. 
Due to the presence of the endpoints of the fundamental strings on the boundary, the local operator should be in a non-trivial 
representation $R$ of a gauge group. 
For example, 
if the boundary field theory would have the gauge group SU$(N)$, then $R$ should be an irreducible component of 
a product of $F_0$ fundamental representations of SU$(N)$. 

There is indeed such an operator in a Chern-Simons-matter theory if it has the gauge group U$(N_1)\times$U$(N_2)$ with the levels 
$k_1,k_2$, respectively, such that 
\begin{equation}
k_1+k_2=F_0. 
   \label{F_0}
\end{equation}

In gauge theories, 
there exists a so-called monopole operator 
\cite{Goddard:1976qe}\cite{Borokhov:2003yu}\cite{Kapustin:2005py} 
which creates a Dirac monopole-like gauge field configuration on $\mathbb{R}^3$. 
Let the monopole be at the origin $O$ of $\mathbb{R}^3$. 
It is necessary to introduce a covering of $\mathbb{R}^3\backslash\{O\}$ 
to construct a non-trivial gauge bundle for the monopole. 
The charge of the monopole is encoded in the transition functions of the bundle. 
One may use a singular gauge transformation to trivialize the transition functions, leaving a Dirac string connecting $O$ and 
the point at infinity. 
If the gauge theory has a Chern-Simons term, then it was shown in 
\cite{Moore:1989yh}\cite{Elitzur:1989nr} that the singular gauge transformation 
also provides an insertion of a Wilson line along the Dirac string\footnote{
More precisely, one starts with cutting out a semi-infinite tube in $\mathbb{R}^3$ which contains the Dirac string. 
Since this procedure creates a boundary, a suitable boundary action should be added to the Chern-Simons action. 
Then, the gauge transformation (which is now regular due to the elimination of the tube) provides a boundary term 
which is equivalent to a Wilson line inserted in the path-integral.
}. 
In the simplest case where the gauge group is U$(1)$ and the monopole has the unit magnetic 
charge, the resulting Wilson line is the one 
describing the propagation of a particle with the electric charge $k$. 

In the theories discussed in \cite{Gaiotto:2009mv}, 
one can construct a so-called di-monopole operator which is a product of two monopole operators for two U$(N)$ factors. 
Let us first consider the case $N_1=N_2=1$ for simplicity. 
By the same argument, it can be shown that 
the di-monopole operator is equivalent to a Wilson line corresponding to a charged particle with 
the electric charges $(k_1,k_2)$. 
It would be natural to relate the electric charges to endpoints of fundamental strings. 
If the levels are chosen such that $k_1,-k_2>0$, then the di-monopole operator should correspond to a state with $k_1$ strings 
with an orientation and $|k_2|$ strings with the opposite orientation. 
A pair of string endpoints with opposite orientations can be contracted with themselves or with bi-fundamental fields, so 
the di-monopole operator would be in a representation which is 
obtained, for example, from a product of $k_1+k_2$ fundamental representation 
of SU$(N_1)$ when $k_1+k_2$ is positive. 
Therefore, under the assumption (\ref{F_0}), one may expect that the di-monopole operator is the dual of the D0-brane with $F_0$ 
fundamental strings attached. 
Since the ground state of a fundamental string stretched between a D2-brane and a D0-brane is bosonic in the trivial flat 
background, it would be natural to expect that this is the case even in the presence of $F_0$, at least when $F_0$ is small. 
This suggests that 
the representation of the Wilson line corresponding to the di-monopole operator is 
$(\rm Sym{\bf N}_1^{k_1},Sym\bar{{\bf N}}_2^{k_2})$ for general $N_1$ and $N_2$. 
Although the Wilson line operator looks non-local, the di-monopole operator is a local operator since the Dirac string is 
unphysical. 
Monopole operators play important roles in Chern-Simons-matter theories 
\cite{Benna:2009xd}\cite{Gustavsson:2009pm}\cite{Kwon:2009ar}\cite{Kim:2009wb}\cite{Berenstein:2009sa}\cite{Kim:2010ac}\cite{Bashkirov:2010kz}\cite{Kim:2009ia}\cite{Samtleben:2010eu}. 

The relation between the Chern-Simons levels and $F_0$ was further elucidated in \cite{Bergman:2010xd}. 
In \cite{Bergman:2010xd}, a brane configuration for the ${\cal N}=0$ version of the theory in \cite{Gaiotto:2009mv} 
was proposed in Type IIB string theory. 
It is basically the same brane configuration used in \cite{Aharony:2008ug}, 
and a non-zero Romans mass is introduced by letting D7-branes pass 
through the system. 
In a T-dual Type IIA string theory, this process corresponds to introducing D8-branes which pass through the system. 
Since D7-branes create a cut for R-R zero form field $C_0$, the effect of the D7-branes still remains even when they are far 
apart. 
The cut for $C_0$ contributes to the Chern-Simons levels, and as a result, the levels which were originally $(k,-k)$ is turned 
into $(k,-k+1)$ when a single D7-brane passes through. 
Repeating this process, or using anti-D7-branes, one can change $k_2$ to an arbitrary value without changing $k_1$.  
 
The explicit gravity solutions were obtained in 
\cite{Tomasiello:2007eq}\cite{Gaiotto:2009mv} which were proposed to be the dual gravity 
backgrounds for the ${\cal N}=0,1$ theories in \cite{Gaiotto:2009mv}. 
Recent researches on solutions of massive Type IIA supergravity are found in 
\cite{Lust:2009zb}\cite{Petrini:2009ur}\cite{Lust:2009mb}\cite{Dall'Agata:2009gv}.

\vspace{1cm}

\section{Planar analysis on Gaiotto-Tomasiello theory} \label{largeN}

\vspace{5mm}

\subsection{Gaiotto-Tomasiello theory} \label{GT}

\vspace{5mm}

In this section, we consider the planar limit of a Chern-Simons-matter theory constructed in \cite{Gaiotto:2009mv}. 
The theory has ${\cal N}=3$ supersymmetry, and the gauge group is 
U$(N_1)\times$U$(N_2)$ with the Chern-Simons levels $k_1$ and $k_2$, respectively. 
It is not necessary to set $k_1+k_2=0$. 
The gauge fields couple to bi-fundamental matter fields. 
The action is 
\begin{eqnarray}
S_{\rm GT} 
&=& S_{\rm CS}+\int d^3x\,\left[ \int d^4\theta\,\left( \mbox{tr}_{2}(A^\dag_ie^{-2V^{(1)}}A_ie^{2V^{(2)}})
    +\mbox{tr}_{1}(B_i^\dag e^{-2V^{(2)}}B_ie^{2V^{(1)}}) \right) \right] \nonumber \\
& & +\int d^3x\, \left[ \int d^2\theta\,\left( \frac{2\pi}{k_1}\mbox{tr}_1(A_iB_iA_jB_j)
    +\frac{2\pi}{k_2}\mbox{tr}_2(B_iA_iB_jA_j) 
    \right) + \mbox{h.c.} \right]
   \label{GT action}
\end{eqnarray}
Here $A_i$ $(i=1,2)$ are ${\cal N}=2$ chiral superfields in the bi-fundamental representation of U$(N_1)\times $U$(N_2)$, 
$B_i$ are 
${\cal N}=2$ chiral superfields in the anti-bi-fundamental representation. 
$V^{(s)}$ $(s=1,2)$ are 
${\cal N}=2$ vector superfields for U$(N_s)$. 
The traces tr$_s$ are the ones for fundamental representations of U$(N_s)$. 
The part $S_{\rm CS}$ in terms of component fields is  
\begin{equation}
S_{\rm CS} 
= \sum_{s=1}^2\ \frac{k_s}{4\pi}\int d^3x\,\mbox{tr}_s\left[ \epsilon^{\mu\nu\rho}\left( A^{(s)}_\mu\partial_\nu A^{(s)}_\rho
    -\frac{2i}3A^{(s)}_\mu A^{(s)}_\nu A^{(s)}_\rho 
    \right) -2i\bar{\lambda}^{(s)}\lambda^{(s)}-2D^{(s)}\sigma^{(s)} \right]. 
\end{equation}
This theory is known to be conformal even quantum mechanically, irrespective of the values of the parameters. 
In the following, we will refer to this theory as Gaiotto-Tomasiello (GT) theory. 
Some properties of ${\cal N}=3$ Chern-Simons-matter theories are summarized in Appendix \ref{N=3 action}. 
See also \cite{Gaiotto:2007qi}. 

\vspace{5mm}

Similarly to ABJ theory, GT theory may not be defined for all possible values of the parameters. 
By an argument similar to the one in \cite{Aharony:2008gk}, it may be suggested that GT theory with parameters satisfying 
\begin{equation}
N_1-N_2 > k_1 > 0
   \label{inconsistent}
\end{equation}
would be inconsistent as a quantum field theory. 
See Appendix \ref{SUSYbreak} for more details. 
In the following, we consider GT theory with parameters which does not satisfy (\ref{inconsistent}), assuming that the theory 
would be consistent. 
Note that one can vary $k_2$ without spoiling the assumed consistency. 

\vspace{5mm}

Since GT theory is superconformal with enough number of supercharges, the localization calculation of 
\cite{Kapustin:2009kz} can apply to 
GT theory as well when it is defined on $S^3$. 
It was shown in \cite{Kapustin:2009kz} that the resulting expression for the partition function is independent of the details of 
the superpotential. 
Since GT theory has exactly the same gauge group and the matter content with ABJ theory, 
the localization formula for the partition function can be obtained 
from that of ABJ theory by simply changing the Chern-Simons levels from $(k,-k)$ to $(k_1,k_2)$. 
The explicit expression is 
\begin{equation}
Z_{\rm GT} 
= \int\prod_{i=1}^{N_1}d\phi_i\prod_{a=1}^{N_2}d\tilde{\phi}_a\,e^{-S(\phi,\tilde{\phi})}, 
   \label{local PF}
\end{equation}
where 
\begin{equation}
e^{-S(\phi,\tilde{\phi})} 
= e^{i\pi\bigl( \sum_ik_1\phi_i^2+k_2\sum_ak_2\tilde{\phi}_a^2 \bigr)}
    \frac{\displaystyle{\prod_{i<j}\sinh^2[\pi(\phi_i-\phi_j)]\prod_{a<b}\sinh^2[\pi(\tilde{\phi}_a-\tilde{\phi}_b)]}}
    {\displaystyle{\prod_{i,a}\cosh^2[\pi(\phi_i-\tilde{\phi}_a)]}}. 
   \label{S_eff}
\end{equation}

\vspace{5mm}

Supersymmetric Wilson loops can be considered in Chern-Simons-matter theories with at least ${\cal N}=2$ supersymmetry 
\cite{Gaiotto:2007qi}. 
The explicit form for GT theory is 
\begin{equation}
W_s[C] = \frac1{N_s}\mbox{tr}_s\mbox{P}\exp\left[ \oint_C d\tau \bigl(i\dot{x}^\mu A^{(s)}_\mu+|\dot{x}|\sigma^{(s)} \bigr) 
 \right], 
   \label{WL}
\end{equation}
where the contour $C$ is a great circle on $S^3$, and $x^\mu=x^\mu(\tau)$ is a parametrization of $C$. 
These operators preserve half of (manifest) ${\cal N}=2$ supersymmetry. 

\vspace{5mm}

Note that the Wilson loops (\ref{WL}) can be also considered in ABJM theory. 
These are actually equivalent to 1/6-BPS Wilson loops constructed in 
\cite{Drukker:2008zx}\cite{Chen:2008bp}\cite{Rey:2008bh} and 
studied further in \cite{Suyama:2009pd}\cite{Marino:2009jd}\cite{Drukker:2010nc}. 
For example, $W_1[C]$ corresponds in ABJM theory to 
\begin{equation}
W[C] 
 = \frac1{N_1}\mbox{tr}_1\mbox{P}\exp\left[ \oint_C d\tau \bigl(i\dot{x}^\mu A^{(1)}_\mu+\frac{2\pi}k|\dot{x}|M_I{}^JY^IY_J \bigr) 
 \right], 
   \label{ABJM WL}
\end{equation}
where $M_I{}^J = \mbox{diag}(+1,+1,-1,-1)$ and $Y^I=(A_i,B^\dag_i)$. 
This correspondence can be checked as follows. 
The action of ABJM theory can be written in terms of ${\cal N}=2$ superfields \cite{Aharony:2008ug}. 
The auxiliary field $\sigma^{(1)}$ can be integrated out by the constraint 
\begin{equation}
\sigma^{(1)} = \frac{2\pi}k(A_iA^\dag_i-B^\dag_iB_i)
   \label{sigma}
\end{equation}
obtained by integrating out another auxiliary field $D^{(1)}$. 
Inserting this into (\ref{WL}), then one obtains the 1/6-BPS Wilson loop (\ref{ABJM WL}). 

The field $\sigma^{(1)}$ is a component of a triplet of SU$(2)_R$ R-symmetry in the ${\cal N}=3$ theory. 
There is a family of supersymmetric Wilson loops which are related to (\ref{ABJM WL}) by R-symmetry rotations. 
Explicitly, 
\begin{equation}
W_1[C,\theta] = \frac1{N_1}\mbox{tr}_1\mbox{P}\exp\left[ \oint_C d\tau \bigl(i\dot{x}^\mu A^{(1)}_\mu
 +|\dot{x}|\theta_i\varphi^i \bigr) \right], 
   \label{UV WL}
\end{equation}
where $\varphi^i$ $(i=1,2,3)$ consist of $\sigma^{(1)}$ 
and an adjoint chiral scalar field\footnote{
The adjoint chiral scalar fields are already integrated out in (\ref{GT action}). 
See Appendix \ref{N=3 action}. 
}, and $\theta_i$ is a unit three-vector. 
This is actually a special case of the operator constructed in \cite{Gaiotto:2007qi}. 
Since $\varphi^i$ are auxiliary fields, they can be integrated out. 
Then (\ref{ABJM WL}) is obtained again with 
\begin{equation}
M_I{}^J = \left[ \theta_i\sigma^i\otimes {\bf 1}_2 \right]_I{}^J, 
\end{equation}
where ${\bf 1}_2$ is the identity $2\times 2$ matrix. 

In a UV completion, for example the worldvolume theory on a brane configuration 
in \cite{Aharony:2008ug}, the fields $\varphi^i$ are dynamical, 
describing motions of the D-branes in transverse 
directions. 
In this UV point of view, (\ref{UV WL}) 
looks very much like the 1/2-BPS Wilson loop in ${\cal N}=4$ super Yang-Mills theory in four 
dimensions. 
This seems to suggest that the information of 
the vector $\theta_i$ which specifies the direction along which a dual string is stretched 
would be encoded in the matrix $M_I{}^J$ in (\ref{ABJM WL}). 

\vspace{5mm}

In GT theory, one can consider both (\ref{WL}) and (\ref{UV WL}). 
Since (\ref{UV WL}) is obtained from (\ref{WL}) by an R-symmetry rotation, let us focus on (\ref{WL}). 
The expectation value of the Wilson loop (\ref{WL}) is expressed in terms of a finite-dimensional integral 
\begin{equation}
\langle W_1[C] \rangle 
 = \frac1{Z_{\rm GT}}\int \prod_{i=1}^{N_1}d\phi_i\prod_{a=1}^{N_2}d\tilde{\phi}_a\,e^{-S(\phi,\tilde{\phi})}
 \frac1{N_1}\sum_{i=1}^{N_1}e^{2\pi \phi_i}
\end{equation}
using the localization \cite{Kapustin:2009kz}.

\vspace{5mm}

\subsection{Planar limit} \label{planar solution}

\vspace{5mm}

We would like to concentrate on the planar limit 
\begin{equation}
k_1,k_2,N_1,N_2 \propto k, \hspace{5mm} k \to \infty. 
    \label{limit}
\end{equation}
In this limit, every observable which can be written in terms of $\phi_i$ and $\tilde{\phi}_a$ is determined by the solution 
of the saddle-point equations derived from $S(\phi,\tilde{\phi})$ in (\ref{S_eff}). 
In terms of rescaled variables 
\begin{equation}
u_i := 2\pi\phi_i, \hspace{5mm} v_a := 2\pi\tilde{\phi}_a, 
\end{equation}
the saddle-point equations are 
\begin{eqnarray}
-\frac{ik_1}{2\pi}u_i 
&=& \sum_{j\ne i}^{N_1}\coth\frac{u_i-u_j}2-\sum_{a=1}^{N_2}\tanh\frac{u_i-v_a}2, 
    \label{GT0-1} \\
-\frac{ik_2}{2\pi}v_a 
&=& \sum_{b\ne a}^{N_2}\coth\frac{v_a-v_b}2-\sum_{i=1}^{N_1}\tanh\frac{v_a-u_i}2. 
    \label{GT0-2}
\end{eqnarray}
As in \cite{Marino:2009jd}, we consider instead the following equations 
\begin{eqnarray}
\frac{k_1}{2\pi}u_i 
&=& \sum_{j\ne i}^{N_1}\coth\frac{u_i-u_j}2+\sum_{a=1}^{N_2}\tanh\frac{u_i-v_a}2, 
   \label{GT0'-1} \\
\frac{k_2}{2\pi}v_a 
&=& \sum_{b\ne a}^{N_2}\coth\frac{v_a-v_b}2+\sum_{i=1}^{N_1}\tanh\frac{v_a-u_i}2, 
   \label{GT0'-2}
\end{eqnarray}
where $k_1,k_2$ are assumed to be positive, 
and the solution of the original equations (\ref{GT0-1})(\ref{GT0-2}) will be obtained via a suitable analytic continuation. 

\vspace{5mm}

Dividing by $k$, one obtains 
\begin{eqnarray}
\kappa_1u_i 
&=& \frac{t_1}{N_1}\sum_{j\ne i}^{N_1}\coth\frac{u_i-u_j}2+\frac{t_2}{N_2}\sum_{a=1}^{N_2}\tanh\frac{u_i-v_a}2, 
    \label{GT-1} \\
\kappa_2v_a 
&=& \frac{t_2}{N_2}\sum_{b\ne a}^{N_2}\coth\frac{v_a-v_b}2+\frac{t_1}{N_1}\sum_{i=1}^{N_1}\tanh\frac{v_a-u_i}2, 
    \label{GT-2}
\end{eqnarray}
where 
\begin{equation}
\kappa_{1,2} := \frac{k_{1,2}}{k}, \hspace{5mm} t_{1,2} := \frac{2\pi N_{1,2}}{k}. 
\end{equation}
The saddle-point equations for ABJ theory correspond to $\kappa_{1,2}=1$. 

Define new variables 
\begin{equation}
z_i := e^{u_i}, \hspace{5mm} w_a := e^{v_a}. 
\end{equation}
In terms of these variables, the equations (\ref{GT-1})(\ref{GT-2}) are written as 
\begin{eqnarray}
\kappa_1\log z_i 
&=& t_1\frac{N_1-1}{N_1}+t_2+\frac{2t_1}{N_1}\sum_{j\ne i}^{N_1}\frac{z_j}{z_i-z_j}
    -\frac{2t_2}{N_2}\sum_{a=1}^{N_2}\frac{w_a}{z_i+w_a}, 
    \label{GT2-1} \\
\kappa_2\log w_a 
&=& t_1+t_2\frac{N_2-1}{N_2}+\frac{2t_2}{N_2}\sum_{b\ne a}^{N_2}\frac{w_b}{w_a-w_b}
    -\frac{2t_1}{N_1}\sum_{i=1}^{N_1}\frac{z_i}{w_a+z_i}. 
    \label{GT2-2}
\end{eqnarray}

Let us assume that all $z_i$ and all $w_a$ are real. 
Define the following resolvent 
\begin{equation}
v(z) := t_1\int dx\,\rho(x)\frac x{z-x}-t_2\int dx\,\tilde{\rho}(x)\frac x{z+x}, 
   \label{v(z)}
\end{equation}
where $\rho(x)$ and $\tilde{\rho}(x)$ are defined formally as 
\begin{equation}
\rho(x) := \frac1{N_1}\sum_{i=1}^{N_1}\delta(x-z_i), \hspace{5mm} \tilde{\rho}(x) := \frac1{N_2}\sum_{a=1}^{N_2}\delta(x-w_a). 
\end{equation}
Note that the overall scale of $v(z)$ depends on the choice of $k$ and therefore irrelevant. 

Assuming that both $z_i$ and $w_a$ form continuous segments on the real axis in the planar limit (\ref{limit}), 
$v(z)$ has two cuts on the $z$-plane. 
Let $[a,b], [c,d] \subset \mathbb{R}$ be the cuts of $v(z)$ corresponding to $-w_a$ and $z_i$, respectively. 
In terms of $v(z)$, the equations (\ref{GT2-1})(\ref{GT2-2}) are written as  
\begin{eqnarray}
\kappa_1\log y -t &=& v(y+i0)+v(y-i0), \hspace{5mm} (c<y<d) 
    \label{GT3-1} \\
\kappa_2\log (-y)-t &=& v(y+i0)+v(y-i0), \hspace{5mm} (a<y<b)
    \label{GT3-2}
\end{eqnarray}
where $t=t_1+t_2$. 
These equations suggest $b<0<c$. 

\vspace{5mm}

It will turn out that the function 
\begin{equation}
F(z,s) := \sqrt{(z-a)(z-b)(z-c)(z-d)}\int_c^ddx\,\frac{f_+(x,s)}{z-x}, 
\end{equation}
where $s$ is a real-valued parameter and 
\begin{equation}
f_\pm(x,s) := \frac{\pm\log(\pm e^{-s}x)}{\sqrt{|(x-a)(x-b)(x-c)(x-d)|}}, 
\end{equation}
is a building block of the solution. 
The phase of the square-root in $F(z,s)$ is chosen such that 
\begin{equation}
\sqrt{(z-a)(z-b)(z-c)(z-d)} = +z^2 + O(z) 
\end{equation}
for $z\to+\infty$. 

Choose $y$ such that $c<y<d$. 
One can show that 
\begin{equation}
F(y\pm i0,s) 
= \pi\log (e^{-s}y)\pm i\sqrt{|(y-a)(y-b)(y-c)(y-d)|}\int_c^ddx\,\mbox{P}\frac{f_+(x,s)}{y-x}. 
\end{equation}
Therefore, $F(z,s)$ satisfies 
\begin{equation}
F(y+i0,s)+F(y-i0,s) = 2\pi\log (e^{-s}y). \hspace{5mm} (c<y<d)
\end{equation}

Next, choose $y$ such that $a<y<b$. 
Then 
\begin{equation}
F(y\pm i0,s) 
= \mp i\sqrt{|(y-a)(y-b)(y-c)(y-d)|}\int_c^ddx\,\mbox{P}\frac{f_+(x,s)}{y-x}, 
\end{equation}
and therefore, 
\begin{equation}
F(y+i0,s)+F(y-i0,s) = 0. \hspace{5mm} (a<y<b)
\end{equation}

Similarly, the other function defined as 
\begin{equation}
G(z,s) := \sqrt{(z-a)(z-b)(z-c)(z-d)}\int_a^bdx\,\frac{f_-(x,s)}{z-x}
\end{equation}
satisfies 
\begin{eqnarray}
G(y+i0,s)+G(y-i0,s) &=& 0, \hspace{25mm} (c<y<d) \\
G(y+i0,s)+G(y-i0,s) &=& 2\pi\log(-e^{-s}y). \hspace{5mm} (a<y<b) 
\end{eqnarray}

It is now easy to check that 
\begin{equation}
v(z) = \frac{\kappa_1}{2\pi}F\left( z,\frac{t}{\kappa_1} \right)+\frac{\kappa_2}{2\pi}G\left( z,\frac t{\kappa_2} \right)
   \label{solution}
\end{equation}
is the solution of the saddle-point equations (\ref{GT3-1})(\ref{GT3-2}). 
Note that a change in $k$ only changes the overall scale, as it should be. 

\vspace{5mm}

The parameters $a,b,c,d$ are constrained by requiring a suitable behavior of $v(z)$ expected from the definition (\ref{v(z)}). 
It is obvious that $v(z)$ must satisfy 
\begin{equation}
v(z) = \left\{ 
\begin{array}{cc}
O(z^{-1}), & (z\to\infty) \\ -t. & (z=0)
\end{array}
\right.
\end{equation}
These conditions imply 
\begin{eqnarray}
\frac{\kappa_1}{2\pi}\int_c^ddx\,f_+\left( x,\frac t{\kappa_1} \right)
 +\frac{\kappa_2}{2\pi}\int_a^bdx\,f_-\left( x,\frac t{\kappa_2} \right) &=& 0, 
   \label{condition1} \\
\frac{\kappa_1}{2\pi}\int_c^ddx\left( x-\frac{a+b+c+d}2 \right)f_+\left( x,\frac t{\kappa_1} \right) & & \nonumber \\
+\frac{\kappa_2}{2\pi}\int_a^bdx\left( x-\frac{a+b+c+d}2 \right)f_-\left( x,\frac t{\kappa_2} \right) &=& 0, 
   \label{condition2} \\
\sqrt{abcd}\left[ \frac{\kappa_1}{2\pi}\int_c^ddx\,x^{-1}f_+\left( x,\frac t{\kappa_1} \right)
 +\frac{\kappa_2}{2\pi}\int_a^bdx\,x^{-1}f_-\left( x,\frac t{\kappa_2} \right) \right] &=& -t. 
   \label{condition3}
\end{eqnarray}
These three conditions are not enough to determine four parameters. 
In fact, this is a situation analogous to the one 
found in multi-cut solutions of one-matrix models (See, for example, \cite{Marino:2004eq}). 
In the context of the one-matrix model, one can impose another condition which forbids the tunneling of the eigenvalues from 
one cut to another. 
In the present context, this kind of condition is not available since the eigenvalues $z_i$ are distinct from $w_a$. 

Noticing that the above conditions are expressed in terms of only $t$, not including 
$t_1$ nor $t_2$ separately, one of the followings 
\begin{equation}
\oint_{C_+}\frac{dz}{2\pi i}\frac{v(z)}z = t_1, \hspace{5mm} \oint_{C_{-}}\frac{dz}{2\pi i}\frac{v(z)}z = t_2, 
   \label{t_1 and t_2}
\end{equation}
which are derived from the definition (\ref{v(z)}), is an independent condition. 
Here $C_+$ is a contour encircles the cut $[c,d]$, and $C_-$ encircles $[a,b]$. 

It is possible to simplify the problem. 
Notice that the original saddle-point equations (\ref{GT0'-1})(\ref{GT0'-2}) are invariant under the simultaneous sign flip of 
$u_i$ and $v_a$. 
This fact implies that the eigenvalue distributions of $z_i$ and $w_a$ should be invariant under the inversion $z\to z^{-1}$ and 
$w\to w^{-1}$. 
We would like to make an ansatz 
\begin{equation}
ab = 1, \hspace{5mm} cd = 1. 
   \label{ansatz}
\end{equation}
However, this ansatz already fixes two parameters in terms of the others, so it should be checked whether the above conditions 
(\ref{condition1})(\ref{condition2})(\ref{condition3}) are compatible with this ansatz. 
Indeed, this is the case. 
Using (\ref{condition1}), the second condition (\ref{condition2}) can be written as 
\begin{equation}
\frac{\kappa_1}{2\pi}\int_{c}^ddx\,xf_+\left( x,\frac t{\kappa_1} \right)
 +\frac{\kappa_2}{2\pi}\int_a^{b}dx\,xf_-\left( x,\frac t{\kappa_2} \right) = 0. 
\end{equation}
Changing the variable to $y=x^{-1}$, this can be written as 
\begin{eqnarray}
 -\frac{\kappa_1}{2\pi}\int_{c}^ddy\,y^{-1}f_+\left( y,\frac t{\kappa_1} \right)
 -\frac{\kappa_2}{2\pi}\int_{a}^{b}dy\,y^{-1}f_-\left( y,\frac t{\kappa_2} \right) 
&=& \frac{t}{\pi}\int_{c}^d\frac{dy}{y\sqrt{|(y-a)(y-b)(y-c)(y-d)|}} \nonumber \\
& & -\frac{t}{\pi}\int_{a}^{b}\frac{dy}{y\sqrt{|(y-a)(y-b)(y-c)(y-d)|}} \nonumber \\
&=& t. 
\end{eqnarray}
This coincides with (\ref{condition3}) under the ansatz (\ref{ansatz}). 

Moreover, the first condition (\ref{condition1}) is trivially satisfied under the ansatz (\ref{ansatz}). 
By the change of the variable $x\to x^{-1}$, one can check that the identities 
\begin{equation}
\int_c^ddx\,f_+(x,0) = 0 = \int_a^bdx\,f_-(x,0) 
\end{equation}
hold. 
Then, the condition (\ref{condition1}) becomes 
\begin{equation}
\int_c^d\,\frac{dx}{\sqrt{|(x-a)(x-b)(x-c)(x-d)|}} - \int_a^b\,\frac{dx}{\sqrt{|(x-a)(x-b)(x-c)(x-d)|}} = 0. 
\end{equation}
It is easy to show that this is an identity. 

Therefore, it is found that there is in fact a single condition for two parameters, verifying the ansatz 
(\ref{ansatz}). 
Still another condition is necessary to determine the parameters completely. 
One can choose it from one of (\ref{t_1 and t_2}). 
Equivalently, one can choose both of (\ref{t_1 and t_2}) as the conditions. 

\vspace{5mm}

In summary, we obtained the planar resolvent 
\begin{equation}
v(z) 
= \sqrt{(z-a)(z-b)(z-c)(z-d)}\left[ \frac{\kappa_1}{2\pi}\int_c^ddx\,\frac{f_+(x,\frac t{\kappa_1})}{z-x} 
    +\frac{\kappa_2}{2\pi}\int_a^bdx\,\frac{f_-(x,\frac t{\kappa_2})}{z-x} \right], 
   \label{result}
\end{equation}
where the parameters are determined by the conditions 
\begin{equation}
ab = 1, \hspace{5mm} cd = 1, \hspace{5mm} 
\oint_{C_+}\frac{dz}{2\pi i}\frac{v(z)}z = t_1, \hspace{5mm} \oint_{C_{-}}\frac{dz}{2\pi i}\frac{v(z)}z = t_2. 
\end{equation}
Recall that we introduced an auxiliary parameter $k$ in order that the dependence of the resolvent on the parameters 
becomes symmetric. 
The above equations imply that the parameters $a,b,c,d$ are independent of the choice of $k$, as it should be the case. 
On the other hand, $v(z)$ is proportional to $k^{-1}$ due to the definition (\ref{v(z)}). 

\vspace{5mm}

It is rather straightforward to show that the solution (\ref{result}) is reduced to the ABJ solution in \cite{Marino:2009jd} when 
$\kappa_1=\kappa_2$, or $k_1=k_2=k$. 
Appendix \ref{ABJ} contains the details of the calculations. 

\vspace{5mm}

It was shown in \cite{Marino:2009jd} that the expectation value $\langle W_1[C] \rangle$ of the supersymmetric 
Wilson loop (\ref{WL}) behaves as 
\begin{equation}
\langle W_1[C] \rangle \sim e^{\pi\sqrt{2\lambda}}
   \label{AdSbehavior}
\end{equation}
for large $\lambda$ in ABJM theory, namely, in the case where $k_1=-k_2=k$, $N_1=N_2=N$, and $\lambda=\frac Nk$. 
Therefore, the Wilson loop has the AdS behavior, confirming the conjecture of \cite{Aharony:2008ug}. 
The behavior (\ref{AdSbehavior}) must be preserved if the deviation of $k_1+k_2$ from zero is negligible in the planar limit. 
This situation would correspond to a gravity solution, a perturbation of AdS$_4$ solution by a Romans mass, 
investigated in \cite{Gaiotto:2009yz}. 
In the next subsection, we will consider the opposite limit in which $k_1+k_2$ goes to infinity. 
In the limit, we will find that the AdS behavior (\ref{AdSbehavior}) disappears.

\vspace{5mm}

\subsection{Decoupling limit} \label{flavored}

\vspace{5mm}

It is interesting to consider the limit $|k_2|\to\infty$ in GT theory. 
If this limit is taken in the perturbation theory, it is obvious that the gauge fields for U$(N_2)$ decouple, and this 
U$(N_2)$ gauge symmetry becomes a global symmetry. 
As a result, a single (anti-)bi-fundamental field becomes $N_2$ (anti-)fundamental fields for U$(N_1)$ gauge group. 
This is simply because one of the 't Hooft coupling, $\frac{N_2}{k_2}$ goes to zero in the limit. 
This is true even non-perturbatively which can be checked by examining the localization formula (\ref{local PF}). 
The limit $|k_2|\to\infty$ can be easily investigated in terms of the stationary phase method. 
Namely, all $\tilde{\phi}_a$ are forced to be zero in the limit, and the resulting partition function coincides with the one for 
${\cal N}=3$ Chern-Simons-matter theory with a single gauge group U$(N_1)$ coupled to $N_2$ fundamental matters and $N_2$ 
anti-fundamental matters\footnote{
More precisely, the partition function approaches that of the ${\cal N}=3$ theory with (anti-)fundamental matters times 
a decoupled system of $\tilde{\phi}_a$ in the limit. 
The decoupled system can be neglected as long as one considers observables independent of $\tilde{\phi}_a$. 
}. 

\vspace{5mm}

A similar limit can be considered in the planar limit discussed in the previous subsection. 
One can take 
\begin{equation}
k_1,k_2,N_1,N_2 \propto k, \hspace{5mm} k \to +\infty, \hspace{5mm} \frac{k}{k_2}\to0. 
\end{equation}
In this limit, the Gaussian terms in (\ref{S_eff}) are 
\begin{equation}
i\pi\left[ \frac{k_1}{N_1}\cdot N_1\sum_{i=1}^{N_1}\phi_i^2+\frac{k_2}{N_2}\cdot N_2\sum_{a=1}^{N_2}\tilde{\phi}_a^2 \right]. 
\end{equation}
As long as $N_1$ and $N_2$ are the same order, and $\frac{N_1}{k_1}$ and $\frac{N_2}{k_2}$ 
are the same order, these two terms are comparable. 
In the limit 
\begin{equation}
\frac{N_2}{k_2} = \frac{N_2}k\frac{k}{k_2} \to 0,
\end{equation} 
the second term dominates, and therefore $\tilde{\phi}_a$ are set to zero. 

\vspace{5mm}

The planar solution of the limit theory may be investigated in terms of the planar resolvent (\ref{result}) of GT theory. 
However, it is far easier to investigate directly the saddle-point equation of the limit theory. 
We would like to solve 
\begin{equation}
u_i+t_2\tanh\frac{u_i}2 = \frac{t_1}{N_1}\sum_{j\ne i}^N\coth\frac{u_i-u_j}2, 
   \label{saddle}
\end{equation}
where we choose $k=k_1$. 
The result of the limit theory will be obtained by continuing $k_1\to-ik_1$ analytically. 

Define the resolvent 
\begin{equation}
v(z) := t_1\int dx\,\rho(x)\frac x{z-x}. 
\end{equation}
In terms of $v(z)$, the equation (\ref{saddle}) can be written as 
\begin{equation}
\log (e^{-t_1}x)+t_2\frac{x-1}{x+1} = v(x+i0)+v(x-i0).  
    \label{saddle2}
\end{equation}
We assume that the solution of (\ref{saddle2}) has a single cut $[a,b]$ with $a>0$. 
It is easy to check that 
\begin{equation}
v(z) = v_0(z)+v_f(z)
\end{equation}
satisfies the saddle-point equation (\ref{saddle2}), where 
\begin{eqnarray}
v_0(z) 
&=& \frac1\pi\int_a^bdx\,\frac1{z-x}\frac{\sqrt{(z-a)(z-b)}}{\sqrt{|(x-a)(x-b)|}}\frac{\log(e^{-t_1}x)}2 \nonumber \\
&=& \frac12\log\left[ \frac{e^{-t_1}}{2\sqrt{ab}+a+b}\left( z+\sqrt{ab}-\sqrt{(z-a)(z-b)} \right)^2 \right], \\
v_f(z) 
&=& \frac1\pi\int_a^bdx\,\frac1{z-x}\frac{\sqrt{(z-a)(z-b)}}{\sqrt{|(x-a)(x-b)|}}\frac{t_2}2\frac{x-1}{x+1} \nonumber \\
&=& \frac{t_2}2\frac{z-1}{z+1}-\frac{t_2}{\sqrt{(1+a)(1+b)}}\frac{\sqrt{(z-a)(z-b)}}{z+1}. 
\end{eqnarray}

The parameters $a$ and $b$ are determined by imposing 
\begin{equation}
v(0) = -t_1, \hspace{5mm} v(\infty) = 0. 
\end{equation}
The condition $v(0)=-t_1$ implies 
\begin{equation}
\frac12\log\left[ \frac{e^{-t_1}(2\sqrt{ab})^2}{2\sqrt{ab}+a+b} \right] 
 - \frac{t_2}2+\frac{t_2\sqrt{ab}}{\sqrt{(1+a)(1+b)}} = -t_1. 
\end{equation}
The other condition $v(\infty)=0$ implies 
\begin{equation}
\frac12\log\left[ \frac{e^{-t_1}}{2\sqrt{ab}+a+b}\left( \sqrt{ab}+\frac{a+b}2 \right)^2 \right]
 + \frac{t_2}2-\frac{t_2}{\sqrt{(1+a)(1+b)}} = 0. 
\end{equation}

As in the case with a finite $k_2$, these two equations are consistent with the ansatz $ab=1$. 
Under this ansatz, the above two conditions become equivalent. 
The remaining equation to be solved is then 
\begin{equation}
\log\frac4{2+a+a^{-1}}+t_2\sqrt{\frac4{2+a+a^{-1}}} = t_2-t_1. 
\end{equation}

\vspace{5mm}

For simplicity, let us focus on the case $t_1=t_2$, that is, $N_1=N_2$. 
The parameter $a$ can be obtained by 
\begin{equation}
a = e^{2u}, \hspace{5mm} 2\cosh u\cdot\log\cosh u = t_1. 
\end{equation}
The expectation value of the Wilson loop is 
\begin{eqnarray}
\langle e^{2\pi \phi} \rangle 
&=& \frac1{t_1}\lim_{z\to\infty}z\,v(z) \nonumber \\
&=& \frac1{t_1}\sinh^2u+\cosh u-1. 
\end{eqnarray}

Let us consider the limit $t_1\to+i\infty$. 
Let $\cosh u=re^{i\theta}$. 
Then 
\begin{eqnarray}
\cos\theta\cdot r\log r-\sin\theta\cdot r\theta &=& 0, \\
\sin\theta\cdot r\log r+\cos\theta\cdot r\theta &=& \frac{|t_1|}2 
\end{eqnarray}
are satisfied. 
These imply 
\begin{equation}
\log r = \theta\tan\theta, \hspace{5mm} \frac{r\log r}{\sin\theta} = \frac{|t_1|}2. 
\end{equation}
The limit $|t_1|\to\infty$ thus corresponds to $r\to\infty$ with $\theta\to\frac{\pi}2-0$. 
Therefore, 
\begin{equation}
|\cosh u| = r = o(t_1). 
\end{equation}
This then implies that 
\begin{equation}
\langle e^{2\pi\phi} \rangle = o(t_1). 
\end{equation}
We found that the exponential behavior observed in the case of small $k_1+k_2$ 
is modified to a behavior which is at most power-like. 
This is a similar phenomenon whose existence was suggested in \cite{Rey:2010ry} for ${\cal N}=2$ superconformal gauge theories 
in four dimensions.

\vspace{1cm}

\section{Discussion} \label{discuss}

\vspace{5mm}

We have considered the planar limit of Gaiotto-Tomasiello theory. 
The planar resolvent (\ref{result}) was obtained in terms of contour integrals. 
The large 't Hooft coupling behavior of supersymmetric Wilson loops was discussed for GT theory and for a limit theory in which 
one of the gauge fields $A^{(2)}_\mu$ decoupled. 
We found that the Wilson loop has the AdS behavior if $k_1+k_2$ is small, but the AdS behavior disappears in the limit 
$k_2\to\infty$. 
This phenomenon is quite similar to the one observed in \cite{Rey:2010ry}. 
In this paper it was shown explicitly in the 
planar limit. 

It does not seem to be easy to understand the dependence on $k_2$ of the behavior of the Wilson loop 
in the gravity dual point of view. 
One reason is that the gravity solutions of massive Type IIA theory are quite complicated. 
Another, possibly more crucial, reason is that the large $k_2$ limit corresponds to a limit in which a field strength $F_0$ 
becomes large. 
Since $F_0$ behaves as a cosmological constant, a large value of $F_0$ would make the curvature of the background geometry 
large, implying that a supergravity approximation would not be valid any more. 
One may argue that the non-exponential behavior found in subsection \ref{flavored} would be due to the fact that the radius of 
the dual AdS$_4$ space might become small, and therefore, the minimal surface in AdS$_4$ would not be dominant. 
To justify this claim, one needs to find a reliable technique to analyze Type IIA string theory in such a background. 

It would be necessary to elaborate on the planar solution in more details. 
Since our solution (\ref{result}) 
is not written in terms of familiar functions, it looks difficult to extract information on the theory. 
However, although implicit, the planar resolvent is completely specified, so whether a necessary piece of information can be 
extracted from (\ref{result}) is a technical issue. 
It would be very interesting if there would exist a variant of the technique used in \cite{Marino:2009jd} which is applicable 
to GT theory. 
Such a technique then may enable one to make it 
possible to show an interpolation of the Wilson loop from a finite $k_2$ to infinite $k_2$. 
Although this limit can be taken rather straightforwardly at the level of the partition function (\ref{local PF}), 
it would be plausible 
to analyze the limit entirely in terms of the planar resolvent (\ref{result}). 
Note that there would not be any singular behavior as $k_2$ goes to infinity. 
Provided that GT theory with a chosen parameters $N_1,N_2,k_1,k_2$ is well-defined, the theory should be kept well-defined 
in the process of increasing $k_2$, since this is just a process of decreasing a coupling constant. 
One might wonder whether there would be a possibility that the discreteness of $k_2$ might change the conclusion. 
However, as far as the planar limit is concerned, there does not seem to be an indication that the discreteness plays a role 
for large $k_2$ (something different may happen for small $k_2$ which corresponds to a strong 't Hooft coupling). 
It would be also interesting to check the behavior of the free energy in the large 't Hooft coupling limit, as in 
\cite{Drukker:2010nc}, 
so as to find the number of the degrees of freedom of GT theory in the limit. 

It would be interesting to extend our analysis to other Chern-Simons-matter theories. 
One interesting theory would be a deformation of ${\cal N}=4$ quiver theory of $\hat{A}_3$-type \cite{Hosomichi:2008jd}. 
This theory may have two distinct Chern-Simons levels, as in GT theory. 
It would be interesting if, by taking one of the levels to be infinity, 
two of the four U$(N)$ gauge fields decouple, leaving Gaiotto-Witten theory 
\cite{Gaiotto:2008sd} coupled to a large number of fundamental matters. 
Comparison of the behavior of Wilson loops in the quiver theory, Gaiotto-Witten theory and 
its deformation by adding fundamental matters might provide an insight into the effects 
of a large number of fundamental matters, or a role of the Veneziano limit, in AdS/CFT correspondence. 

\vspace{2cm}

{\bf \Large Acknowledgements}

\vspace{5mm}

We would like to thank Soo-Jong Rey and Jaesung Park for valuable discussions and comments. 
This work was supported by the BK21 program of the Ministry of Education, Science and Technology, 
National Science Foundation of Korea Grants R01-2008-000-10656-0, 
2005-084-C00003, 2009-008-0372 and EU-FP Marie Curie Research 
\& Training Network HPRN-CT-2006-035863 (2009-06318).

\vspace{2cm}

\appendix

{\bf \LARGE Appendices}

\vspace{1cm}

\section{${\cal N}=3$ Chern-Simons-matter theory} \label{N=3 action}

\vspace{5mm}

To construct the Lagrangian of an ${\cal N}=3$ Chern-Simons-matter theory, it is convenient to employ the ${\cal N}=1$ superspace 
formalism in four dimensions accompanied by the dimensional reduction. 
For notations and conventions, we basically follow \cite{Wess:1992cp}. 

\vspace{5mm}

Let $\psi$ denote $\psi_\alpha$, and $\epsilon^{\alpha\beta}$ etc. will be written explicitly. 
The three-dimensional gamma matrices are 
\begin{equation}
\gamma^0 := i\sigma^2, \hspace{5mm} \gamma^1 := \sigma^3, \hspace{5mm} \gamma^2 := -\sigma^1. 
\end{equation}
The adjoint $\bar{\psi}$ is defined as usual: 
\begin{equation}
\bar{\psi} := \psi^\dag\gamma^0. 
\end{equation}

In the dimensional reduction, we ignore all $x^2$-dependence, and then redefine the indices suitably. 
For example, let $v_m$ be a four-dimensional vector field. 
After the dimensional reduction, we denote $v_2$ by $\sigma$, and the other components $v_0,v_1,v_3$ by $A_\mu$. 
The ${\cal N}=2$ supersymmetry transformations for the vector multiplet fields are 
\begin{eqnarray}
\delta A_\mu 
&=& -i\bar{\lambda}\gamma_\mu\xi+i\bar{\xi}\gamma_\mu\lambda, \\ 
\delta \sigma 
&=& \bar{\lambda}\xi-\bar{\xi}\lambda, \\
\delta \lambda 
&=& -\frac12\gamma^{\mu\nu}\xi F_{\mu\nu}-i\gamma^\mu\xi D_\mu\sigma+i\xi D, \\
\delta D 
&=& D_\mu\bar{\lambda}\gamma^\mu\xi+\bar{\xi}\gamma^\mu D_\mu\lambda+[\sigma,\bar{\lambda}]\xi+\bar{\xi}[\sigma,\lambda], 
\end{eqnarray}
where $D_\mu\lambda=\partial_\mu\lambda-i[A_\mu,\lambda]$ etc. 
It is straightforward to check that the action 
\begin{equation}
S_{\rm CS} 
= \frac{k}{4\pi}\int d^3x\,\mbox{tr}\left[ \epsilon^{\mu\nu\rho}\left( A_\mu\partial_\nu A_\rho-\frac{2i}3A_\mu A_\nu A_\rho 
 \right) -2i\bar{\lambda}\lambda-2D\sigma \right]
   \label{CS}
\end{equation}
is invariant under the above transformation. 
It is possible to write this action explicitly in terms of the vector superfield $V$ \cite{Gaiotto:2007qi}. 

\vspace{5mm}

To construct an ${\cal N}=3$ Lagrangian, we introduce an adjoint chiral superfield $\Phi$, and two chiral superfields $Q$ and 
$\tilde{Q}$ in $R$ and $\bar{R}$ representations, respectively, of the gauge group. 
This is nothing but the field content of an ${\cal N}=2$ gauge theory with matters in four dimensions. 
The following Lagrangian 
\begin{equation}
S_{\rm matter} 
= \int d^3x \left[ 
 \int d^4\theta\, \left[ Q^\dag e^{-2V}Q+\tilde{Q}e^{2V}\tilde{Q}^\dag \right] - \int d^2\theta \sqrt{2}\tilde{Q}\Phi Q
 - \int d^2\bar{\theta} \sqrt{2}Q^\dag\Phi^\dag \tilde{Q}^\dag \right]
   \label{matter}
\end{equation}
in three dimensions possesses SU$(2)_R\times $SU$(2)_{\rm rot}$ 
global symmetry if it is supplemented by ordinary kinetic terms of $V$ and $\Phi$. 
Here SU$(2)_R$ is the R-symmetry of the ${\cal N}=2$ theory in four dimensions. 
On the other hand, 
SU$(2)_{\rm rot}$ has a geometric origin if the action is regarded as a part of the dimensional reduction of an ${\cal N}=1$ 
gauge theory with matters in six dimensions. 
Namely, it is the rotational symmetry in the three dimensional subspace which is 
reduced in the dimensional reduction. 

Obviously, SU$(2)_R$ symmetry does not commute with ${\cal N}=2$ supersymmetry in three dimensions. 
If there would exist an action which includes (\ref{CS}) and (\ref{matter}), and which is invariant under the SU$(2)_R$ symmetry, 
then the action should have a larger supersymmetry. 
However, the action $S_{\rm CS}+S_{\rm matter}$ does not have the SU$(2)_R$ symmetry. 
It is, for example, due to the constraint 
\begin{equation}
\sigma = \frac{2\pi}k(qq^\dag-\tilde{q}^\dag\tilde{q}), 
\end{equation}
where the indices for $R$ and $\bar{R}$ are not contracted in the right-hand side. 
Notice that the right-hand side is not a singlet of $SU(2)_R$. 
In fact, it is, say, the third component of a triplet of SU$(2)_R$. 

This is not the end of the story. 
Since $\sigma$ is also a component of a triplet of SU$(2)_{\rm rot}$, it is possible that a diagonal SU$(2)_d$ subgroup of 
SU$(2)_R\times $SU$(2)_{\rm rot}$ can be a symmetry of a theory. 
It can be shown that 
\begin{equation}
S_{{\cal N}=3} = S_{\rm CS}+S_{\rm matter}-\int d^3x\left[ \int d^2\theta\, \frac k{4\pi}\mbox{tr}\,\Phi^2+
 \int d^2\bar{\theta}\, \frac k{4\pi}\mbox{tr}\,(\Phi^\dag)^2 \right] 
   \label{N=3}
\end{equation}
has such a symmetry. 
Therefore, this theory has at least ${\cal N}=3$ supersymmetry. 
It was shown in \cite{Gaiotto:2008sd} that a special choice of the gauge group and the field content has to be made for 
a Chern-Simons-matter theory 
to possess ${\cal N}=4$ supersymmetry. 
Therefore, generically, this theory has only ${\cal N}=3$ supersymmetry. 
The ${\cal N}=3$ supersymmetry transformation was explicitly written in \cite{Kao:1992ig}. 

Note that the action (\ref{N=3}) is completely determined by the requirement of ${\cal N}=3$ supersymmetry except for the value 
of $k$. 
Due to the invariance for large gauge transformations, $k$ must be an integer. 
This implies that $k$ is not renormalized beyond one-loop since the loop expansion parameter of the theory is $k^{-1}$. 
It is known that the presence of ${\cal N}=2$ supersymmetry ensures the non-renormalization theorem of $k$ 
\cite{Kao:1995gf}. 
Therefore, since (\ref{N=3}) is classically superconformal after integrating $\Phi$ out, it is also 
superconformal quantum mechanically.

\vspace{1cm}

\section{ABJ slice} \label{ABJ}

\vspace{5mm}

In subsection \ref{planar solution}, we obtained the planar resolvent of GT theory 
\begin{equation}
v(z) = \frac{\kappa_1}{2\pi}F\left( z,\frac{t}{\kappa_1} \right)+\frac{\kappa_2}{2\pi}G\left( z,\frac t{\kappa_2} \right), 
\end{equation}
where 
\begin{eqnarray}
F(z,s) &=& \int_c^ddx\,\frac{\log (e^{-s}x)}{z-x}\frac{\sqrt{(z-a)(z-b)(z-c)(z-d)}}{\sqrt{|(x-a)(x-b)(x-c)(x-d)|}}, \\
G(z,s) &=& -\int_a^bdx\frac{\log (-e^{-s}x)}{z-x}\frac{\sqrt{(z-a)(z-b)(z-c)(z-d)}}{\sqrt{|(x-a)(b-x)(c-x)(x-d)|}}. 
\end{eqnarray}
We assume $ab=1$ and $cd=1$ whose validity is shown in subsection \ref{planar solution}. 
The functions $F(z,s), G(z,s)$ can be written as follows. 
\begin{eqnarray}
\frac1\pi F(z,s) 
&=& \oint_{C_+}\frac{dx}{2\pi i}\,\frac{\log(e^{-s}x)}{z-x}\frac{\sqrt{(z-a)(z-b)(z-c)(z-d)}}{\sqrt{(x-a)(x-b)(x-c)(x-d)}}, \\
\frac1\pi G(z,s) 
&=& \oint_{C_-}\frac{dx}{2\pi i}\,\frac{\log(-e^{-s}x)}{z-x}\frac{\sqrt{(z-a)(z-b)(z-c)(z-d)}}{\sqrt{(x-a)(x-b)(x-c)(x-d)}}, 
\end{eqnarray}
where the contour $C_+$ encircles the cut $[c,d]$ and $C_-$ encircles $[a,b]$. 

\vspace{5mm}

To check the validity of the solution, consider the ABJ slice $\kappa_1=\kappa_2=1$ and see whether the solution in 
\cite{Marino:2009jd} is reproduced. 
On this slice, 
\begin{eqnarray}
2v(z) 
&=& \oint_{C}\frac{dx}{2\pi i}\,\frac{\log(e^{-s}x)}{z-x}\frac{\sqrt{(z-a)(z-b)(z-c)(z-d)}}{\sqrt{(x-a)(x-b)(x-c)(x-d)}} 
    \nonumber \\
& & +\pi i\oint_{C_-}\frac{dx}{2\pi i}\,\frac{1}{z-x}
    \frac{\sqrt{(z-a)(z-b)(z-c)(z-d)}}{\sqrt{(x-a)(x-b)(x-c)(x-d)}}, 
   \label{from sign}
\end{eqnarray}
where $C=C_+\cup C_-$ and $s=t/2$. 
Note that to define $G(z,s)$ without ambiguity, we should 
choose $\log(-e^{-s}x)=\log(e^{-s+\pi i}x)$, and the cut $[a,b]$ should be in fact 
$[a-i0,b-i0]$. 
The branch cut of the logarithm is chosen to be the negative real axis. 

Consider the first contour integral. 
By deforming the contour, 
\begin{eqnarray}
& & \oint_{C}\frac{dx}{2\pi i}\,\frac{\log(e^{-s}x)}{z-x}\frac{\sqrt{(z-a)(z-b)(z-c)(z-d)}}{\sqrt{(x-a)(x-b)(x-c)(x-d)}} 
    \nonumber \\
&=& -\oint_z\frac{dx}{2\pi i}\,\frac{\log(e^{-s}x)}{z-x}\frac{\sqrt{(z-a)(z-b)(z-c)(z-d)}}{\sqrt{(x-a)(x-b)(x-c)(x-d)}} 
    \nonumber \\
& & -\oint_{C_{\log}}\frac{dx}{2\pi i}\,\frac{\log(e^{-s}x)}{z-x}\frac{\sqrt{(z-a)(z-b)(z-c)(z-d)}}{\sqrt{(x-a)(x-b)(x-c)(x-d)}}, 
\end{eqnarray}
where $C_{\log}$ encircles the negative real axis. 
The integral around $z$ provides $\log(e^{-s}z)$. 
The integral along $C_{\log}$ can be written, using the ordinary trick, 
\begin{eqnarray}
& & -\oint_{C_{\log}}\frac{dx}{2\pi i}\,\frac{\log(e^{-s}x)}{z-x}\frac{\sqrt{(z-a)(z-b)(z-c)(z-d)}}{\sqrt{(x-a)(x-b)(x-c)(x-d)}}
    \nonumber \\
&=& \int_{-\infty}^0\frac{dx}{z-x}\frac{\sqrt{(z-a)(z-b)(z-c)(z-d)}}{\sqrt{(x-a)(x-b)(x-c)(x-d)}}. 
\end{eqnarray}
It turns out that the second term in (\ref{from sign}) cancels a part of the integral. 
Indeed, 
\begin{eqnarray}
& & \int_{a}^b\frac{dx}{z-x}\frac{\sqrt{(z-a)(z-b)(z-c)(z-d)}}{\sqrt{(x-a)(x-b)(x-c)(x-d)}} \nonumber \\
&=& \int_{a+i0}^{b+i0}\frac{dx}{z-x}\frac{\sqrt{(z-a)(z-b)(z-c)(z-d)}}{\sqrt{(x-a)(x-b)(x-c)(x-d)}} \nonumber \\
&=& -\frac12\oint_{C_-}\frac{dx}{z-x}\frac{\sqrt{(z-a)(z-b)(z-c)(z-d)}}{\sqrt{(x-a)(x-b)(x-c)(x-d)}}. 
\end{eqnarray}

As a result, 
\begin{eqnarray}
2v(z) 
&=& \log(e^{-s}z)+\left( \int_{-\infty}^a+\int_b^0 \right)\frac{dx}{z-x}
    \frac{\sqrt{(z-a)(z-b)(z-c)(z-d)}}{\sqrt{(x-a)(x-b)(x-c)(x-d)}} \nonumber \\
&=& \log(e^{-s}z)-\frac1z\int_{-\infty}^adx\,\frac{x^2-1}{x^2-(z+z^{-1})x+1}
    \frac{\sqrt{(z-a)(z-b)(z-c)(z-d)}}{\sqrt{(x-a)(x-b)(x-c)(x-d)}}. 
\end{eqnarray}
The integral in the second line can be performed explicitly by the use of the formula 
\begin{equation}
\int dx\,\frac{x^2-1}{x^2-cx+1}\frac1{\sqrt{(x^2-ax+1)(x^2-bx+1)}} = 2\int \frac{dt}{(c-a)t^2-(c-b)}
\end{equation}
where $t=\sqrt{(x^2-bx+1)/(x^2-ax+1)}$. 
The integration in the right-hand side can be performed easily. 
Finally, we obtain 
\begin{equation}
v(z) = \log\left[ \frac{e^{-\frac t2}}{\sqrt{(c+d)-(a+b)}}\left( \sqrt{(z-a)(z-b)}-\sqrt{(x-c)(z-d)} \right) \right].
\end{equation}
This resolvent is related to $\omega(z)$ in \cite{Marino:2009jd} as 
\begin{equation}
\omega(z) = t+2v(z). 
\end{equation}
Using the relation \cite{Marino:2009jd} 
\begin{equation}
\sqrt{(c+d)-(a+b)} = \sqrt{2\xi} = 2e^{\frac t2}, 
\end{equation}
we reproduce Eq.(3.18) of \cite{Marino:2009jd}.

\vspace{1cm}

\section{Consistency of GT theory} \label{SUSYbreak}

\vspace{5mm}

GT theory is reduced to ABJ theory by setting $k_1=-k_2=k$. 
It was argued in \cite{Aharony:2008gk} that ABJ theory would be inconsistent as a quantum field theory if $|N_1-N_2|$ 
is larger than the Chern-Simons level $k$. 
This can be explained as follows. 

Recall that ABJ theory has a brane configuration whose worldvolume theory flows to ABJ theory in the IR limit. 
The brane configuration consists of an NS5-brane, a $(1,k)$5-brane, $N$ D3-branes wrapping a compact direction, and $l$ segments 
of D3-branes suspended between the NS5-brane and the $(1,k)$5-brane. 
The wrapped D3-branes can move freely, but the D3-brane segments cannot since they must be fixed at a position so as to minimize 
their energy due to the tension. 
In a point of the moduli space for which the wrapped D3-branes are far apart from the D3-brane segments, the low energy effective 
theory of the worldvolume degrees of freedom is the sum of the worldvolume theory on the wrapped D3-branes and the 
worldvolume theory on the D3-brane segments. 
The latter theory is ${\cal N}=3$ pure Yang-Mills Chern-Simons theory with the gauge group SU$(l)$ and the level $k$. 
It was shown in \cite{Kitao:1998mf}\cite{Bergman:1999na}\cite{Ohta:1999iv}
that the supersymmetry is broken in this theory if $l>k$. 

One can consider a similar situation in ABJ theory directly. 
The separation of the wrapped D3-branes corresponds to introducing a vev of a scalar field, like \cite{Mukhi:2008ux}. 
The low energy effective theory around the non-trivial vev includes ${\cal N}=3$ pure Chern-Simons theory 
which always has a supersymmetric vacuum because of the absence of quantum corrections. 
Therefore, there would exist a discrepancy in the pattern of supersymmetry breaking 
between the result from the brane picture and the field theory result. 
It was argued in \cite{Aharony:2008gk} that this may suggest an inconsistency of ABJ theory for the case $l>k$. 

\vspace{5mm}

Although the corresponding brane configuration for GT theory is not known, a similar analysis can be performed. 
For definiteness, we assume $N_1>N_2$ and $k_1>0$. 
Consider GT theory expanded around the vacuum 
\begin{equation}
\langle A_1 \rangle = \left[ 
\begin{array}{c}
v\cdot {\bf I}_{N_2} \\ 0
\end{array}
\right], \hspace{5mm} \langle A_2 \rangle, \langle B_i \rangle = 0, 
\end{equation}
where ${\bf I}_{N_2}$ is $N_2\times N_2$ identity matrix. 
We consider the large $v$ limit. 

The kinetic term of $A_1$ provides mass terms for gauge fields. 
Let $A_\mu^{(1)}$ be decomposed as 
\begin{equation}
A_\mu^{(1)} = \left[ 
\begin{array}{cc}
a_\mu & W^\dag_\mu \\ W_\mu & C_\mu
\end{array}
\right], 
\end{equation}
where $a_\mu$ are $N_2\times N_2$ matrix-valued, and $C_\mu$ are $(N_1-N_2)\times (N_1-N_2)$ matrix-valued. 
Then, the covariant derivative of $A_1$ becomes  
\begin{equation}
D_\mu A_1 = -iv\left[ 
\begin{array}{c}
a_\mu - A_\mu^{(2)} \\ W_\mu
\end{array}
\right] + \cdots. 
\end{equation}
Since $W_\mu$ acquire a large mass in the large $v$ limit, $W_\mu$ should be set to zero at low energy. 
Then, one can show that the Chern-Simons term for $A_\mu^{(1)}$ becomes a sum of Chern-Simons terms for $a_\mu$ and $C_\mu$ 
with a common level $k_1$. 

If all the matter fields coupled to $C_\mu$ become infinitely heavy in the large $v$ limit, then $C_\mu$ would be gauge fields in 
${\cal N}=3$ pure Chern-Simons theory at low energy. 
According to \cite{Aharony:2008gk}, the appearance of this theory as a part of 
the low energy effective theory may suggest that GT theory 
might be inconsistent if $N_1-N_2>k_1$. 
To see whether this is the case, it is enough to check the masses of bi-fundamental fermions, because of the presence of 
supersymmetry. 
Let $\psi_i$ be fermions in the superfield $A_i$ and let $\tilde{\psi}_i$ be fermions in $B_i$. 
We denote them by 
\begin{equation}
\psi_i = \left[ 
\begin{array}{c}
\chi_i \\ \eta_i
\end{array}
\right], \hspace{5mm} \tilde{\psi}_i = \left[ 
\begin{array}{cc}
\tilde{\chi}_i & \tilde{\eta}_i
\end{array}
\right], 
\end{equation}
where $\chi_i, \tilde{\chi}_i$ are $N_2\times N_2$ matrix-valued. 
The gauge fields $C_\mu$ couple to $\eta_i$ and $\tilde{\eta}_i$. 

The superpotential in (\ref{GT action}) does not provide mass terms for $\eta_i$ nor $\tilde{\eta}_i$. 
The mass terms of those fermions come from the Yukawa couplings 
\begin{equation}
\bar{\psi}_i\sigma^{(1)}\psi_i-\bar{\psi}_i\psi_i\sigma^{(2)}
 +\bar{\tilde{\psi}}_i\sigma^{(2)}\tilde{\psi}_i-\bar{\tilde{\psi}}_i\tilde{\psi}_i\sigma^{(1)},  
\end{equation}
where $\sigma^{(1)},\sigma^{(2)}$ are determined by the component scalars $A_i$ and $B_i$ 
(see, for example, (\ref{sigma}) for the case $k_1=-k_2=k$). 
These terms provide masses of order $v^2$ for all components of $\eta_i$ and $\tilde{\eta}_i$. 
Therefore, they are integrated out at low energy.


\vspace{2cm}


\begin{thebibliography}{99}

\bibitem{Maldacena:1997re}
  J.~M.~Maldacena,
  ``The large N limit of superconformal field theories and supergravity,''
  Adv.\ Theor.\ Math.\ Phys.\  {\bf 2}, 231 (1998)
  [Int.\ J.\ Theor.\ Phys.\  {\bf 38}, 1113 (1999)]
  [arXiv:hep-th/9711200].

\bibitem{Rey:1998ik}
  S.~J.~Rey and J.~T.~Yee,
  ``Macroscopic strings as heavy quarks in large N gauge theory and  anti-de
  Sitter supergravity,''
  Eur.\ Phys.\ J.\  C {\bf 22}, 379 (2001)
  [arXiv:hep-th/9803001].

\bibitem{Maldacena:1998im}
  J.~M.~Maldacena,
  ``Wilson loops in large N field theories,''
  Phys.\ Rev.\ Lett.\  {\bf 80}, 4859 (1998)
  [arXiv:hep-th/9803002].

\bibitem{Rey:1998bq}
  S.~J.~Rey, S.~Theisen and J.~T.~Yee,
  ``Wilson-Polyakov loop at finite temperature in large N gauge theory and
  anti-de Sitter supergravity,''
  Nucl.\ Phys.\  B {\bf 527}, 171 (1998)
  [arXiv:hep-th/9803135].

\bibitem{Erickson:2000af}
  J.~K.~Erickson, G.~W.~Semenoff and K.~Zarembo,
  ``Wilson loops in N = 4 supersymmetric Yang-Mills theory,''
  Nucl.\ Phys.\  B {\bf 582}, 155 (2000)
  [arXiv:hep-th/0003055].

\bibitem{Pestun:2007rz}
  V.~Pestun,
  ``Localization of gauge theory on a four-sphere and supersymmetric Wilson
  loops,''
  arXiv:0712.2824 [hep-th].

\bibitem{Aharony:2008ug}
  O.~Aharony, O.~Bergman, D.~L.~Jafferis and J.~Maldacena,
  ``N=6 superconformal Chern-Simons-matter theories, M2-branes and their
  gravity duals,''
  JHEP {\bf 0810}, 091 (2008)
  [arXiv:0806.1218 [hep-th]].

\bibitem{Kapustin:2009kz}
  A.~Kapustin, B.~Willett and I.~Yaakov,
  ``Exact Results for Wilson Loops in Superconformal Chern-Simons Theories with Matter,''
  JHEP {\bf 1003}, 089 (2010)
  [arXiv:0909.4559 [hep-th]].

\bibitem{Drukker:2008zx}
  N.~Drukker, J.~Plefka and D.~Young,
  ``Wilson loops in 3-dimensional N=6 supersymmetric Chern-Simons Theory and
  their string theory duals,''
  JHEP {\bf 0811}, 019 (2008)
  [arXiv:0809.2787 [hep-th]].

\bibitem{Chen:2008bp}
  B.~Chen and J.~B.~Wu,
  ``Supersymmetric Wilson Loops in N=6 Super Chern-Simons-matter theory,''
  Nucl.\ Phys.\  B {\bf 825}, 38 (2010)
  [arXiv:0809.2863 [hep-th]].

\bibitem{Rey:2008bh}
  S.~J.~Rey, T.~Suyama and S.~Yamaguchi,
  ``Wilson Loops in Superconformal Chern-Simons Theory and Fundamental Strings
  in Anti-de Sitter Supergravity Dual,''
  JHEP {\bf 0903}, 127 (2009)
  [arXiv:0809.3786 [hep-th]].

\bibitem{Drukker:2009hy}
  N.~Drukker and D.~Trancanelli,
  ``A supermatrix model for N=6 super Chern-Simons-matter theory,''
  JHEP {\bf 1002}, 058 (2010)
  [arXiv:0912.3006 [hep-th]].

\bibitem{Suyama:2009pd}
  T.~Suyama,
  ``On Large N Solution of ABJM Theory,''
  Nucl.\ Phys.\  B {\bf 834}, 50 (2010)
  [arXiv:0912.1084 [hep-th]].

\bibitem{Marino:2009jd}
  M.~Marino and P.~Putrov,
  ``Exact Results in ABJM Theory from Topological Strings,''
  JHEP {\bf 1006}, 011 (2010)
  [arXiv:0912.3074 [hep-th]].

\bibitem{Drukker:2010nc}
  N.~Drukker, M.~Marino and P.~Putrov,
  ``From weak to strong coupling in ABJM theory,''
  arXiv:1007.3837 [hep-th].

\bibitem{Rey:2010ry}
  S.~J.~Rey and T.~Suyama,
  ``Exact Results and Holography of Wilson Loops in N=2 Superconformal (Quiver) Gauge Theories,''
  arXiv:1001.0016 [hep-th].

\bibitem{Gadde:2009dj}
  A.~Gadde, E.~Pomoni and L.~Rastelli,
  ``The Veneziano Limit of N=2 Superconformal QCD: Towards the String Dual of N=2 SU($N_c$) SYM with $N_f =2 N_c$,''
  arXiv:0912.4918 [hep-th].

\bibitem{Gaiotto:2009tk}
  D.~Gaiotto and D.~L.~Jafferis,
  ``Notes on adding D6 branes wrapping RP3 in AdS4 x CP3,''
  arXiv:0903.2175 [hep-th].

\bibitem{Hikida:2009tp}
  Y.~Hikida, W.~Li and T.~Takayanagi,
  ``ABJM with Flavors and FQHE,''
  JHEP {\bf 0907}, 065 (2009)
  [arXiv:0903.2194 [hep-th]].

\bibitem{Ammon:2009wc}
  M.~Ammon, J.~Erdmenger, R.~Meyer, A.~O'Bannon and T.~Wrase,
  ``Adding Flavor to AdS4/CFT3,''
  JHEP {\bf 0911}, 125 (2009)
  [arXiv:0909.3845 [hep-th]].

\bibitem{Jafferis:2009th}
  D.~L.~Jafferis,
  ``Quantum corrections to N=2 Chern-Simons theories with flavor and their AdS4 duals,''
  arXiv:0911.4324 [hep-th].

\bibitem{Gaiotto:2009mv}
  D.~Gaiotto and A.~Tomasiello,
  ``The gauge dual of Romans mass,''
  JHEP {\bf 1001}, 015 (2010)
  [arXiv:0901.0969 [hep-th]].

\bibitem{Aharony:2008gk}
  O.~Aharony, O.~Bergman and D.~L.~Jafferis,
  ``Fractional M2-branes,''
  JHEP {\bf 0811}, 043 (2008)
  [arXiv:0807.4924 [hep-th]].

\bibitem{Bergshoeff:1996cy}
  E.~Bergshoeff and M.~De Roo,
  ``D-branes and T-duality,''
  Phys.\ Lett.\  B {\bf 380}, 265 (1996)
  [arXiv:hep-th/9603123].

\bibitem{Green:1996bh}
  M.~B.~Green, C.~M.~Hull and P.~K.~Townsend,
  ``D-Brane Wess-Zumino Actions, T-Duality and the Cosmological Constant,''
  Phys.\ Lett.\  B {\bf 382}, 65 (1996)
  [arXiv:hep-th/9604119].

\bibitem{Goddard:1976qe}
  P.~Goddard, J.~Nuyts and D.~I.~Olive,
  ``Gauge Theories And Magnetic Charge,''
  Nucl.\ Phys.\  B {\bf 125}, 1 (1977).

\bibitem{Borokhov:2003yu}
  V.~Borokhov,
  ``Monopole operators in three-dimensional N = 4 SYM and mirror symmetry,''
  JHEP {\bf 0403}, 008 (2004)
  [arXiv:hep-th/0310254].

\bibitem{Kapustin:2005py}
  A.~Kapustin,
  ``Wilson-'t Hooft operators in four-dimensional gauge theories and S-duality,''
  Phys.\ Rev.\  D {\bf 74}, 025005 (2006)
  [arXiv:hep-th/0501015].

\bibitem{Moore:1989yh}
  G.~W.~Moore and N.~Seiberg,
  ``Taming the Conformal Zoo,''
  Phys.\ Lett.\  B {\bf 220}, 422 (1989).

\bibitem{Elitzur:1989nr}
  S.~Elitzur, G.~W.~Moore, A.~Schwimmer and N.~Seiberg,
  ``Remarks On The Canonical Quantization Of The Chern-Simons-Witten Theory,''
  Nucl.\ Phys.\  B {\bf 326}, 108 (1989).

\bibitem{Benna:2009xd}
  M.~K.~Benna, I.~R.~Klebanov and T.~Klose,
  ``Charges of Monopole Operators in Chern-Simons Yang-Mills Theory,''
  JHEP {\bf 1001}, 110 (2010)
  [arXiv:0906.3008 [hep-th]].

\bibitem{Gustavsson:2009pm}
  A.~Gustavsson and S.~J.~Rey,
  ``Enhanced N=8 Supersymmetry of ABJM Theory on R(8) and R(8)/Z(2),''
  arXiv:0906.3568 [hep-th].

\bibitem{Kwon:2009ar}
  O.~K.~Kwon, P.~Oh and J.~Sohn,
  ``Notes on Supersymmetry Enhancement of ABJM Theory,''
  JHEP {\bf 0908}, 093 (2009)
  [arXiv:0906.4333 [hep-th]].

\bibitem{Kim:2009wb}
  S.~Kim,
  ``The complete superconformal index for N=6 Chern-Simons theory,''
  Nucl.\ Phys.\  B {\bf 821}, 241 (2009)
  [arXiv:0903.4172 [hep-th]].

\bibitem{Berenstein:2009sa}
  D.~Berenstein and J.~Park,
  ``The BPS spectrum of monopole operators in ABJM: towards a field theory description of the giant torus,''
  JHEP {\bf 1006}, 073 (2010)
  [arXiv:0906.3817 [hep-th]].

\bibitem{Kim:2010ac}
  H.~C.~Kim and S.~Kim,
  ``Semi-classical monopole operators in Chern-Simons-matter theories,''
  arXiv:1007.4560 [hep-th].

\bibitem{Bashkirov:2010kz}
  D.~Bashkirov and A.~Kapustin,
  ``Supersymmetry enhancement by monopole operators,''
  arXiv:1007.4861 [hep-th].

\bibitem{Bergman:2010xd}
  O.~Bergman and G.~Lifschytz,
  ``Branes and massive IIA duals of 3d CFT's,''
  JHEP {\bf 1004}, 114 (2010)
  [arXiv:1001.0394 [hep-th]].

\bibitem{Tomasiello:2007eq}
  A.~Tomasiello,
  ``New string vacua from twistor spaces,''
  Phys.\ Rev.\  D {\bf 78}, 046007 (2008)
  [arXiv:0712.1396 [hep-th]].

\bibitem{Lust:2009zb}
  D.~Lust and D.~Tsimpis,
  ``Classes of AdS4 type IIA/IIB compactifications with SU(3)xSU(3) structure,''
  JHEP {\bf 0904}, 111 (2009)
  [arXiv:0901.4474 [hep-th]].

\bibitem{Petrini:2009ur}
  M.~Petrini and A.~Zaffaroni,
  ``N=2 solutions of massive type IIA and their Chern-Simons duals,''
  JHEP {\bf 0909}, 107 (2009)
  [arXiv:0904.4915 [hep-th]].

\bibitem{Lust:2009mb}
  D.~Lust and D.~Tsimpis,
  ``New supersymmetric AdS4 type II vacua,''
  JHEP {\bf 0909}, 098 (2009)
  [arXiv:0906.2561 [hep-th]].

\bibitem{Dall'Agata:2009gv}
  G.~Dall'Agata, G.~Villadoro and F.~Zwirner,
  ``Type-IIA flux compactifications and N=4 gauged supergravities,''
  JHEP {\bf 0908}, 018 (2009)
  [arXiv:0906.0370 [hep-th]].

\bibitem{Gaiotto:2007qi}
  D.~Gaiotto and X.~Yin,
  ``Notes on superconformal Chern-Simons-matter theories,''
  JHEP {\bf 0708}, 056 (2007)
  [arXiv:0704.3740 [hep-th]].


\bibitem{Marino:2004eq}
  M.~Marino,
  ``Les Houches lectures on matrix models and topological strings,''
  arXiv:hep-th/0410165.

\bibitem{Gaiotto:2009yz}
  D.~Gaiotto and A.~Tomasiello,
  ``Perturbing gauge/gravity duals by a Romans mass,''
  J.\ Phys.\ A  {\bf 42}, 465205 (2009)
  [arXiv:0904.3959 [hep-th]].

\bibitem{Hosomichi:2008jd}
  K.~Hosomichi, K.~M.~Lee, S.~Lee, S.~Lee and J.~Park,
  ``N=4 Superconformal Chern-Simons Theories with Hyper and Twisted Hyper
  JHEP {\bf 0807}, 091 (2008)
  [arXiv:0805.3662 [hep-th]].


\bibitem{Gaiotto:2008sd}
  D.~Gaiotto and E.~Witten,
  ``Janus Configurations, Chern-Simons Couplings, And The Theta-Angle in N=4 Super Yang-Mills Theory,''
  JHEP {\bf 1006}, 097 (2010)
  [arXiv:0804.2907 [hep-th]].

\bibitem{Wess:1992cp}
  J.~Wess and J.~Bagger,
  ``Supersymmetry and supergravity,''
{\it  Princeton, USA: Univ. Pr. (1992) 259 p}

\bibitem{Kao:1992ig}
  H.~C.~Kao and K.~M.~Lee,
  ``Selfdual Chern-Simons systems with an N=3 extended supersymmetry,''
  Phys.\ Rev.\  D {\bf 46}, 4691 (1992)
  [arXiv:hep-th/9205115].

\bibitem{Kao:1995gf}
  H.~C.~Kao, K.~M.~Lee and T.~Lee,
  ``The Chern-Simons coefficient in supersymmetric Yang-Mills Chern-Simons theories,''
  Phys.\ Lett.\  B {\bf 373}, 94 (1996)
  [arXiv:hep-th/9506170].

\bibitem{Halmagyi:2003ze}
  N.~Halmagyi and V.~Yasnov,
  ``The spectral curve of the lens space matrix model,''
  JHEP {\bf 0911}, 104 (2009)
  [arXiv:hep-th/0311117].

\bibitem{Kitao:1998mf}
  T.~Kitao, K.~Ohta and N.~Ohta,
  ``Three-dimensional gauge dynamics from brane configurations with (p,q)-fivebrane,''
  Nucl.\ Phys.\  B {\bf 539}, 79 (1999)
  [arXiv:hep-th/9808111].

\bibitem{Bergman:1999na}
  O.~Bergman, A.~Hanany, A.~Karch and B.~Kol,
  ``Branes and supersymmetry breaking in 3D gauge theories,''
  JHEP {\bf 9910}, 036 (1999)
  [arXiv:hep-th/9908075].

\bibitem{Ohta:1999iv}
  K.~Ohta,
  ``Supersymmetric index and s-rule for type IIB branes,''
  JHEP {\bf 9910}, 006 (1999)
  [arXiv:hep-th/9908120].

\bibitem{Mukhi:2008ux}
  S.~Mukhi and C.~Papageorgakis,
  ``M2 to D2,''
  JHEP {\bf 0805}, 085 (2008)
  [arXiv:0803.3218 [hep-th]].

\bibitem{Kim:2009ia}
  S.~Kim and K.~Madhu,
  ``Aspects of monopole operators in N=6 Chern-Simons theory,''
  JHEP {\bf 0912}, 018 (2009)
  [arXiv:0906.4751 [hep-th]].

\bibitem{Hohenegger:2009as}
  S.~Hohenegger and I.~Kirsch,
  ``A note on the holography of Chern-Simons matter theories with flavour,''
  JHEP {\bf 0904}, 129 (2009)
  [arXiv:0903.1730 [hep-th]].

\bibitem{Fujita:2009xz}
  M.~Fujita and T.~S.~Tai,
  ``Eschenburg space as gravity dual of flavored N=4 Chern-Simons-matter theory,''
  JHEP {\bf 0909}, 062 (2009)
  [arXiv:0906.0253 [hep-th]].

\bibitem{Benini:2009qs}
  F.~Benini, C.~Closset and S.~Cremonesi,
  ``Chiral flavors and M2-branes at toric CY4 singularities,''
  JHEP {\bf 1002}, 036 (2010)
  [arXiv:0911.4127 [hep-th]].

\bibitem{Fujita:2009kw}
  M.~Fujita, W.~Li, S.~Ryu and T.~Takayanagi,
  ``Fractional Quantum Hall Effect via Holography: Chern-Simons, Edge States,
  and Hierarchy,''
  JHEP {\bf 0906}, 066 (2009)
  [arXiv:0901.0924 [hep-th]].

\bibitem{Samtleben:2010eu}
  H.~Samtleben and R.~Wimmer,
  ``N=6 Superspace Constraints, SUSY Enhancement and Monopole Operators,''
  arXiv:1008.2739 [hep-th].
play 
\end{thebibliography}
\end{document}